\documentclass[10pt,letterpaper,twocolumn]{article} 

\usepackage{ol2} 

\usepackage{amsmath}

\usepackage{graphicx}
\usepackage{epstopdf}

\begin{document}

\twocolumn[ 

\title{Multi-scale Optics for Enhanced Light Collection from a Point Source}

\author{Rachel Noek,$^{1,*}$ Caleb Knoernschild,$^1$ Justin Migacz,$^1$ Taehyun Kim,$^1$ Peter Maunz,$^1$ True Merrill,$^2$ Harley Hayden,$^3$ C.S. Pai,$^3$ and Jungsang Kim$^1$}
\address{$^1$Fitzpatrick Institute for Photonics, Electrical and Computer Engineering Department, Duke University, Durham, N.C. 27708, USA\\
$^2$School of Chemistry and Biochemistry, Georgia Institute of Technology, Atlanta, Georgia 30332, USA\\
$^3$Georgia Tech Research Institute, Georgia Institute of Technology, Atlanta, Georgia 30332, USA\\$^*$Corresponding author: rachel.noek@duke.edu}

\begin{abstract}
High efficiency collection of photons emitted by a point source over a wide field-of-view (FoV) is crucial for many applications. Multi-scale optics offer improved light collection by utilizing small optical components placed close to the optical source, while maintaining a wide FoV provided by conventional imaging optics. In this work, we demonstrate collection efficiency of 26\% of photons emitted by a point-like source using a micromirror fabricated in silicon with no significant decrease in collection efficiency over a 10 mm object space.
\end{abstract}

]

Efficient collection of photons emitted by a point source requires an optical system with high numerical aperture (NA). It is difficult to design an optical system featuring a high NA over a wide field-of-view (FoV) using cost-effective conventional refractive optical elements. Lens systems with NA=0.85 and a FoV of over 25 mm have been realized for lithography applications \cite{MatsuyamaSPIE2006}. However, such optical systems utilize a large number of lens elements and suffer from optical loss, complexity, size, weight and cost. In conventional applications, the NA of the collection optics is limited to about 0.5, corresponding to a 7\% collection efficiency of photons emitted from a point source. Use of reflective optics, like curved mirrors, opens up the possibility of dramatically enhancing the photon collection efficiency. Recent experiments and proposals using trapped ions demonstrate the benefit of adequate reflective optical elements for imaging \cite{ShuJPB2009}, state detection, and ion-photon coupling applications \cite{MaiwaldNaturePhysics2009,LuoPhysik2009}. While these approaches can dramatically increase the photon collection efficiency, macroscopic reflectors suffer from large geometric aberrations, which need to be corrected in order to distinguish light from multiple point sources. In this work, we employ a multi-scale optical design \cite{Brady2009} to increase the photon collection efficiency from a single point source, which can be extended to high efficiency collection from an array of point sources. This design uses a single conventional objective lens and places a high NA micromirror behind each point source to allow for high efficiency collection from each point source. The ability to image the point sources in a continuous FoV is sacrificed in exchange for high efficiency collection from each point source in a discontinuous FoV. In this way, dramatic improvements are possible in integration time and data acquisition speed for applications where image resolution is determined by the light excitation source and not by the collection optics, including determination of the internal state of a single atom \cite{NagourneyPRL1986,BergquistPRL1986,SauterPRL1986}, confocal laser scanning microscopy \cite{DenkScience1990}, and confocal Raman microspectroscopy \cite{PuppelsNature1990}. The collection efficiency from each point source is determined by the high NA of the micromirror, while the number of point source-micromirror combinations that can be measured simultaneously is determined by the FoV of the macroscopic imaging system. 
\begin{figure}[!ht]
\centering
\includegraphics[width=3.25in]{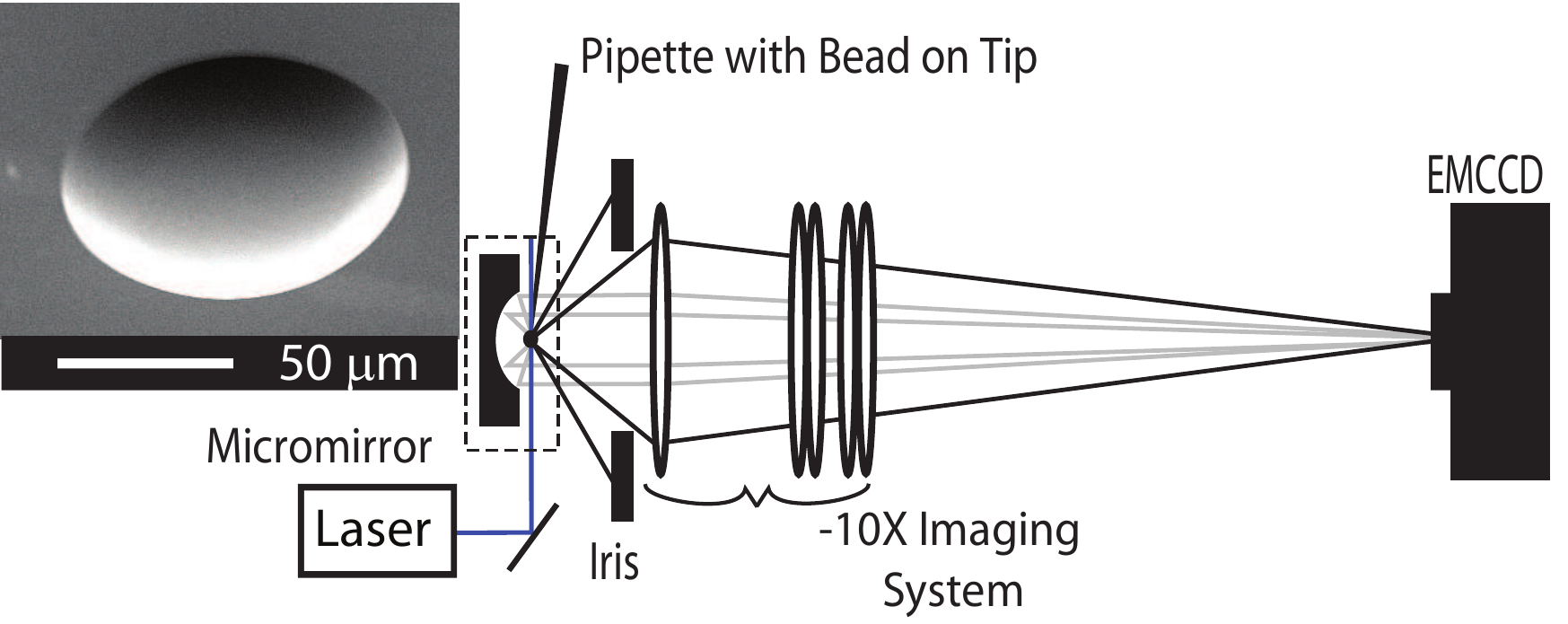}
\caption{Experimental setup with a macroscopic $F/2$ imaging system and a large NA micromirror (inset), using a fluorescent microbead as a point source.}
\label{figure:system}
\end{figure}

In our system design [Fig. \ref{figure:system}], a conventional macroscopic lens system images a point source onto a detector. The imaging system is designed and simulated with optical design software (Zemax$^{\texttt{\textregistered}}$) and features a 2 inch aperture and $F/2$, a magnification of $-10$, and an effective focal length of 90 mm. A micromirror placed behind the point source is used to collect a large fraction of the emitted photons and redirect them toward the macroscopic imaging system. The point source is placed at the focal point of the micromirror, which  collimates the portion of the light emitted within its NA. The imaging system relays the reflected light from the micromirror onto its image plane, where it is detected with an electron-multiplying charge-coupled device (EMCCD, Andor$^{\text{TM}}$ iXon$^{\text{EM}}$) with a $512\times512$ array of 16 $\mu$m square pixels \cite{Kim2009}.

The micromirror is fabricated on a $<$100$>$ silicon wafer coated with a silicon nitride mask layer. An array of mask openings ranging in diameter from 5 to 80 $\mu$m is etched in the nitride mask using reactive ion etching. A 1:8:1 hydrofluoric, nitric and acetic acid (HNA) bath is used \cite{Trupke2005} to isotropically etch the exposed silicon for 15 minutes. The acid ratios determine the etch rate and final surface roughness of the silicon \cite{Robbins1960}, and we choose the combination that optimizes surface smoothness ($<$0.5 nm RMS) and etch rate (4.5 $\mu$m/min). Using an unagitated etch, we can control the center location, depth, and radius of curvature of the micromirrrors to within $\pm$2 $\mu$m of target values over a large wafer area (4 in.), which is crucial for mirror array fabrication. After the etch, the wafer is temporarily coated with thick ($\sim$15 $\mu$m) SiO$_2$ to protect the mirrors while the wafer undergoes surface grinding and chemical mechanical polishing to reduce the mirror sag. A scanning electron microscope (SEM) image of a typical mirror is shown in the inset of Fig. \ref{figure:system}. A mask opening of 60 $\mu$m was used to fabricate the micromirror used in the measurement, which has a 100 $\mu$m radius of curvature measured at the bottom, 90 $\mu$m opening diameter and a 15.6 $\mu$m sag. The sample is sputtered with aluminum, with a reflectance of $80\%$ at 532 nm.

\begin{figure}[!ht]
\centering
\includegraphics[width=3.25in]{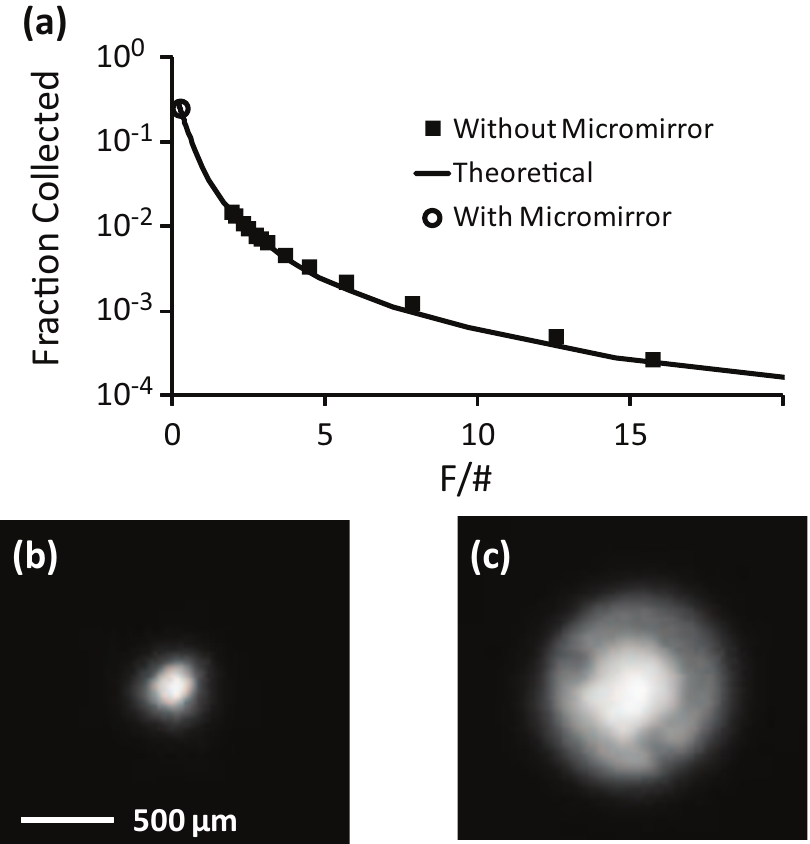}
\caption{(a) Plot of system $F/\#$ with (circle) and without (squares) the micromirror compared to theory (Eq. (1)). Images of the bead (b) without and (c) with the micromirror.}
\label{figure:beadpair}
\end{figure}
In order to demonstrate the enhancement in light collection, a 15 $\mu$m diameter polystyrene fluorescent microbead (FluoSpheres 465 nm, Invitrogen) is used as a point source. The bead is attached to a sharp glass pipette coated with UV curing epoxy. The pipette is mounted on an $xyz$ micromanipulator with 0.2 $\mu$m resolution. The bead is optically pumped with a 407 nm diode laser and has a broad emission spectrum centered at 465 nm. A 500 nm bandpass filter (FB-500-10-1, Thorlabs) blocks the scattered pump photons from entering the detector.  

The collection efficiency of the macroscopic imaging system is calibrated by placing an iris in front of the first lens to control the collection solid angle of the system. The collection efficiency of the photons emitted by a point source using an imaging system is related to its $F/\#$ by [solid line in Fig. \ref{figure:beadpair}(a)]
\begin{equation}
\eta \propto \frac{1}{2}[1-\text{cos}(\text{arctan}(\frac{1}{2F/\#}))].
\label{eq:eqn1}
\end{equation}
We compare the number of background-corrected photon counts detected from the microbead fluorescence at different iris openings with Eq. (\ref{eq:eqn1}) to determine the collection efficiency of the imaging system [Fig. \ref{figure:beadpair}(a)]. When the iris is fully open, 1.5\% of the light emitted from the bead is collected with the imaging optics, corresponding to $F/2$ of the imaging system [image shown in Fig. \ref{figure:beadpair}(b)]. The collection efficiency is corrected for the transmission of the imaging system (not including the bandpass filter), which is about 96\% at 532 nm in our setup.

With the calibrated imaging optics in place, the custom fabricated micromirror is added to the system. The micromirror is attached to a kinematic mirror mount on an independent $xyz$ stage with 1 $\mu$m resolution. The micromirror is aligned so that the microbead is at the mirror focal point in order to collimate the incident light toward the detector and maximize the collection efficiency. After background correction, the counts are further corrected for the shadowing due to the pipette tip holding the microbead ($\sim$2\% of total counts), to estimate the fraction of light collected. When the micromirror is added, the collection efficiency dramatically improves to 26$\pm$3\% corresponding to 34\% solid angle coverage when finite reflectance of the micromirror is accounted for [circle in Fig. \ref{figure:beadpair}(a)]. Figure \ref{figure:beadpair}(c) shows the image of the point source-micromirror combination at the EMCCD plane, showing noticeable enhancement in detected photon counts.

\begin{figure}
\centering
\includegraphics[width=3.25in]{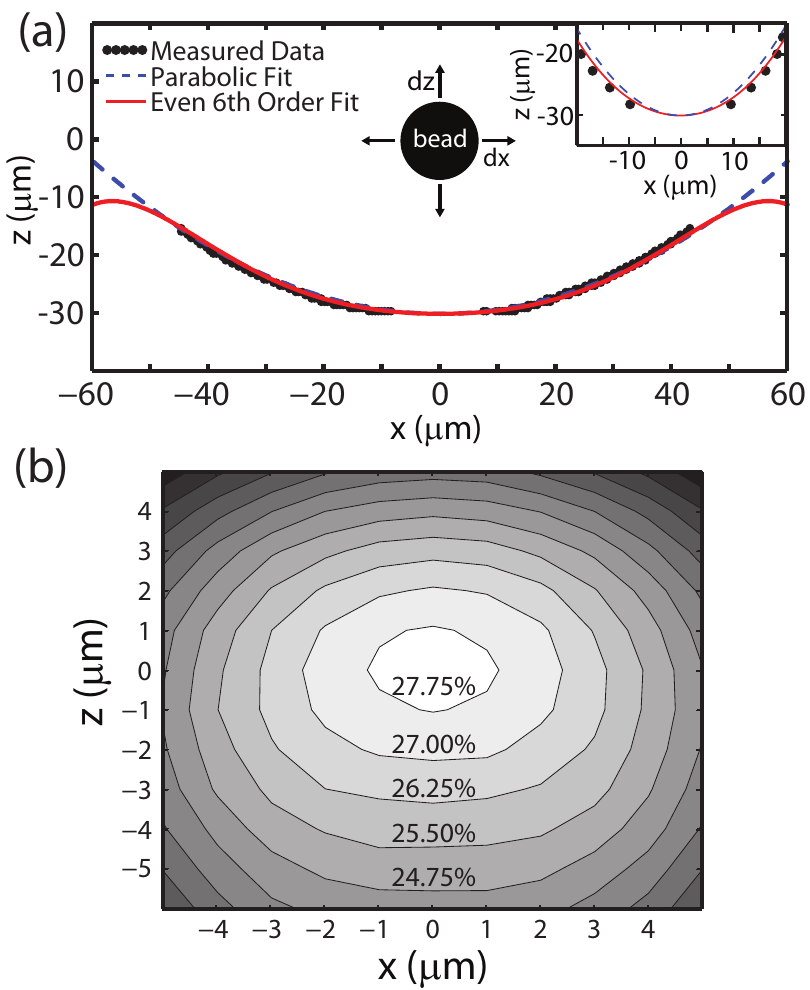}
\caption{(a) Cross sectional profile of a similar micromirror measured by SEM (dotted line), 6th order even polynomial fit (solid line), and K=-1 parabolic fit (dashed line). The inset shows the data near the bottom of the micromirror. (b) The simulated collection efficiency as the bead is moved away from the focal point of the mirror along the $x$ and $z$ directions.}
\label{figure:robust}
\end{figure}
Figure \ref{figure:robust}(a) shows a fit to the profile of the mirror from a cross-sectional SEM  image of a micromirror similar to the one used in the experiment. The image data is fit to a sixth-order even polynomial (inset shows the micromirror surface contrasted with a parabolic surface) and compared to the simulated collection efficiency of several aspherical surfaces (conic constants K = -1, 0, 1). Our fitted model yields $<$1\% lower collection efficiency than ideal conic surfaces. Furthermore, the collection scheme can tolerate up to $\pm$4 $\mu$m of misalignment of the point source from the focal point of the micromirror in both $x$ and $z$ directions with modest reduction ($<$3.25\%) in collection efficiency [Fig. \ref{figure:robust}(b)], which is important for practical applications in atomic fluorescence and scanning confocal microscopy.

\begin{figure}[!ht]
\centering
\includegraphics[width=3.25in]{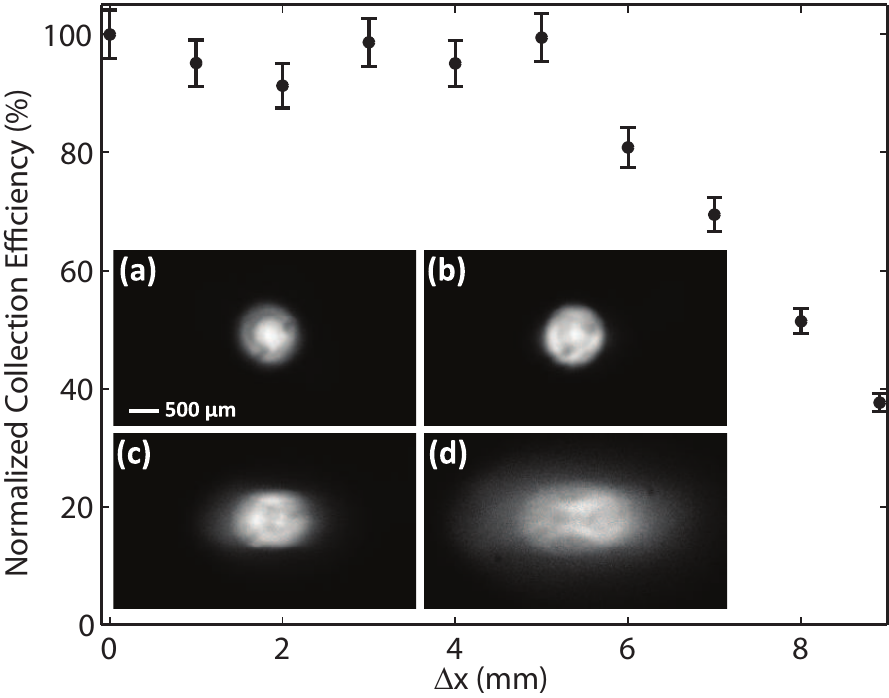}
\caption{Change in measured collection efficiency with bead-micromirror pair position. Inset images show distortion of bead-micromirror pair with $\Delta x$ of (a) 0 mm, (b) 3 mm, (c) 5 mm, and (d) 7 mm from the optical axis.}
\label{figure:FoV}
\end{figure}

Simultaneous detection of multiple point source-micromirror combinations is simulated by horizontally translating the imaging system with respect to the point source-micromirror pair and adjusting the EMCCD imager accordingly. The light emitted by two point sources can be distinguished as long as the images of the point source-micromirror pairs do not overlap in the image plane. The insets in Fig. \ref{figure:FoV} show the image of the pair as a function of their offset from the optical axis, $\Delta x$. The image develops noticeable distortions beyond a 5 mm offset due to geometric aberrations in the imaging system. Figure \ref{figure:FoV} shows the collection efficiency for a fixed size of the integration area (corresponding to $200\;\mu$m$\times200\;\mu$m at the object plane) as a function of $\Delta x$, normalized by the value at $\Delta x=0$. The collection efficiency within this integration area did not significantly degrade when $|\Delta x|\leq5$ mm. From this, we anticipate that our system can simultaneously collect light from over $2\times10^3$ distinguishable point sources without significant degradation in collection efficiency or crosstalk, as long as the separation between the point sources allows positioning of a micromirror behind each source. Beyond this offset value, the integration area must be substantially increased to collect most of the light collimated by the micromirror, reducing the density of distinguishable point source-micromirror pairs that can be imaged simultaneously. Use of a smaller micromirror or a macro-scale imaging system with better aberration correction to reduce the blur can lead to simultaneous collection from a larger array of point sources.

We have shown that a multi-scale optical system can be used to dramatically improve the collection efficiency of light from multiple point sources simultaneously. The micromirror could be integrated with ion traps to achieve a factor of 5 enhancement in light collection over the current state-of-the-art of 5\% \cite{KingThesis}.

The authors would like the thank Curtis Volin for the original imaging system design. This work was supported by the Intelligence Advanced Research Projects Activity under Army Research Office contract number W911NF08-1-0315.

\end{document}